\documentclass[runningheads]{llncs}
\usepackage[T1]{fontenc}
\usepackage{graphicx}
\usepackage{array}
\usepackage{booktabs}
\usepackage{tikz}
\usetikzlibrary{arrows.meta,positioning}
\usepackage{hyperref}
\begin{document}
\title{AI Forensics Across White-, Grey-, and Black-Box Access: A Process Model and Research Agenda for Post-Incident Investigation of AI Systems}

\author{Ali Dehghantanha\inst{1}\orcidID{0000-0001-6385-9937} \and
Sajad Homayoun\inst{2}\orcidID{0000-0001-9371-9759}}
\authorrunning{A. Dehghantanha et al.}
\titlerunning{AI Forensics Across Access Regimes}
\institute{Cyber Science Lab, University of Guelph, Canada \and
Cyber Security Group, Dept. of Electronic Systems, Aalborg University, Denmark\\
\email{adehghan@uoguelph.ca}\\
\email{sajadh@es.aau.dk}\\
}
\maketitle              % typeset the header of the contribution

\begin{abstract}
AI systems are increasingly involved in decisions and actions that may later require investigation. When an AI related incident occurs, investigators need to reconstruct what the system did, why it behaved that way, and which part of the system or supply chain contributed to the outcome. Existing work on AI forensics remains fragmented, often focusing on a specific system type, artifact, or analysis technique. This paper argues that investigator access is a useful starting point for organizing the field. We distinguish white box, grey box, and black box access and show how each access level changes what can be collected, preserved, analyzed, and reported. Based on this distinction, we propose a process model matrix for AI forensics across four phases: collection, preservation, analysis, and reporting. We also introduce an order of volatility for AI systems, covering runtime state, context windows, logs, retrieval stores, model artifacts, and training lineage. From this matrix, we derive an access conditioned examination framework and identify open research problems, including black box preservation, model version attestation, uncertainty quantification for surrogate based analysis, and chain of custody for mutable AI artifacts.

\end{abstract}

\section{Introduction}
\label{sec:intro}
    AI systems increasingly make or mediate consequential decisions such as approving credit, triaging patients, moderating speech, and, in their agentic form, taking actions in the world on a user's behalf. As deployment has widened, so has the surface for failure: training data is poisoned, fine-tunes are backdoored, prompts are injected, retrieval stores are corrupted, and otherwise well-behaved systems drift into harmful conduct. When such an incident occurs (e.g. harm, attack, or regulatory breach) someone must reconstruct what the system did, why it did it, and who or what is responsible. That reconstructive task belongs to forensics, but the methods, evidence models, and tooling for applying it to AI systems remain immature relative to the stakes.

    A terminological hazard must be cleared at the outset. In much of the literature, \textit{AI forensics} denotes the use of artificial intelligence to assist digital-forensic tasks such as automating triage, classification, or evidence recovery \cite{Dunsin2024}. In this paper, we use the term in the opposite, and we argue more urgent, sense: the forensic investigation of AI systems as the subjects of inquiry. This usage is not itself new. Baggili and Behzadan~\cite{baggili2019} explicitly proposed AI forensics as a domain in this sense, sketching sub-areas including model authentication, model identification, and model-malware forensics, but in the half-decade since, the domain has not consolidated so much as fragmented.
    
    Subsequent contributions have been valuable but disconnected: forensic taxonomies for agentic systems, model-authenticity and provenance techniques, reanimation tooling for failed models, and practitioner accounts of model examination \cite{sremack2026} each address a slice of the problem, typically bound to a particular system type or to an unstated assumption about how much access the investigator has to the system under examination. Our position is that these efforts are facets of a single discipline and that the axis along which they should be unified is not system type but investigator access. Whether one holds a model's weights and training data, only its deployment logs and traces, or merely its inputs and outputs, changes every forensic phase, for example, what can be collected, what can be preserved, what can be concluded, and with what confidence it can be reported. We therefore organize AI forensics around three access regimes: white box, grey box, and black box access, and map the forensic process onto them.
    
    This paper makes the following contributions.
    \begin{itemize}
        \item We disambiguate and consolidate AI forensics as the investigation of AI systems, situating it against adjacent fields (interpretability, auditing, observability, and safety).
        \item We argue for investigator access as the primary organizing axis and define the white-, grey-, and black-box regimes.
        \item We present a process model matrix that relates forensic phases to access levels, places existing fragmented work into its cells, and exposes empty cells as open problems.
        \item We propose an order of volatility for AI systems that governs collection priority.
        \item We give a practical, access-conditioned examination framework for investigators and law-enforcement practitioners.
        \item We derive a research agenda directly from the matrix's gaps.
    
    \end{itemize}
    
    \paragraph{Motivating incident.}
    To keep the discussion concrete, we carry a single incident throughout the paper. A regulated financial advisory firm deploys a third party agentic LLM assistant that uses a system prompt, a retrieval store, customer interaction history, and a transaction execution tool guarded by policy checks. Over several weeks, the assistant begins to steer clients toward a fraudulent product and eventually triggers an unauthorized transaction. Several causes are plausible: a poisoned fine tuning set, a corrupted retrieval entry, a prompt injection path, a model update, or a weakness in the tool authorization layer. Three parties investigate the same incident from different vantages: the deploying firm in a grey box posture, the model provider in a white box posture, and an external regulator or incident response team in a black box posture. We return to this scenario at each forensic phase to show how the same incident supports different conclusions depending on access.

\section{Scope and Related Work}
\label{sec:scope}

    We scope AI forensics as the post incident, evidentiary, and access constrained investigation of AI systems. These three terms define the boundary of the paper.
    Post-incident distinguishes forensics from the continuous monitoring of observability and the pre-deployment scrutiny of auditing.
    Evidentiary means that findings must withstand an external challenge in court, before a regulator, or in a contractual dispute. This imposes requirements for integrity, reproducibility, and documented custody that go beyond ordinary debugging.
    Access-constrained captures the reality that an investigator rarely commands the whole system; what can be known is bounded by what can be reached. The literature relevant to this scope is real and growing, but it has accreted as disconnected fragments rather than a coherent discipline. This section establishes this claim.
    
    \subsection{Traditional Forensic Process Models Under-Fit}

    The established process models of digital forensics provide useful foundations for the identification, preservation, analysis, and reporting of evidence. The DFRWS roadmap, NIST SP 800 86 and the ISO IEC 27037 and 27043 standards describe parts of this process, including evidence handling, forensic analysis, and incident investigation procedures~\cite{palmer2001roadmap,kent2006guide,iso27037,iso27043}. They already account for problems such as volatility, distributed evidence, and chain of custody. AI systems, however, add evidence classes that these models do not describe in detail, including model weights, checkpoints, training and fine tuning data, prompts, retrieved context, tool call traces, and mutable deployment state. They also make reproducibility harder, since behavior may depend on stochastic decoding, changing retrieval stores, hidden model routing, or provider side updates. The issue is therefore not that traditional forensic models are irrelevant, but that they need AI specific extensions to cover the evidence surface and failure modes of modern AI systems.

    \subsection{A Field Assembled from Fragments}
    AI forensics, in the sense used in this paper, was named by Baggili and Behzadan~\cite{baggili2019}, who proposed it as forensic investigation of AI systems and sketched several early problem areas, including model authentication, model identification and extraction, and model malware forensics.
    This founding taxonomy remains a useful map, but it predates the foundation-model era and is model-centric, saying little about the serving pipelines, retrieval stores, and agentic action traces that dominate contemporary incidents.

    Recent work has begun to fill in specific parts of the problem. For agentic systems, forensic analysis of an autonomous agent has yielded an artifact taxonomy that decomposes an agent into context, configuration, execution, model interaction, and scheduling planes, and identifies the nondeterministic nature of model decision making as a fundamental limit on reconstruction~\cite{openclaw2026}. For model forensics, Schneider and Breitinger frame the foundational attribution question as a forensic problem in its own right: whether an AI system caused a specific event, and if so, what triggered the action~\cite{schneider2023}. In the reconstruction technique, \textit{reanimation} tooling reconstructs a failed model's decision state from recovered artifacts to determine what went wrong, an approach explicitly motivated by the fact that full access to a proprietary system is not guaranteed \cite{aip2024}. In terms of preservation, model-provenance methods such as model-DNA fingerprinting address whether one model derives from another \cite{modeldna2023}. Each of these is a genuine advance; each is also partial, bound either to one system type, one forensic phase, or one access posture, and none situates itself within a process model that the others share.
    That the discipline is broader than any research thread is evidenced from the practitioner side. The demand for forensic analysis of models arises not only from compliance, but also from mergers and acquisitions, restructuring, private-equity diligence, and vendor assurance \cite{sremack2026}. The market is converging on a discipline that the literature has not yet unified.
    
    \subsection{Differentiation from the Closest Prior Work}

    Because the OpenClaw study \cite{openclaw2026} is the closest neighbor to our work, we precisely state the relationship. The OpenClaw study provides a deep artifact level decomposition of a single agentic system investigated from one access posture. Its artifact planes are, in our terms, a detailed expansion of grey box collection and analysis for one system type. Our contribution is different: we hold the forensic phases fixed and vary access across white box, grey box, and black box regimes. This makes the OpenClaw artifact planes one well populated region of a larger map, while leaving other regions, including black box preservation, surrogate based black box analysis, and cross boundary custody, outside its scope. We adopt its caution about nondeterminism and generalize it from a property of agentic systems to a property that affects AI forensics more broadly.

    \subsection{Adjacent Fields}

    AI forensics borrows methods from several mature areas without being reducible to any of them. Interpretability and explainability supply analysis techniques, such as attribution, probing, and feature visualization. However, these techniques are usually designed to explain model behaviour, individual predictions, or feature attribution in general settings, rather than to preserve evidence and reconstruct a specific past incident under custody and access constraints \cite{BarredoArrieta2020}.
    Model auditing shares forensics' rigor but is pre-deployment and prospective rather than post-incident. The observability of MLOps provides much of the raw material (e.g., logs, traces, metrics), but treats it as operational telemetry, not as evidence subject to custody and challenge \cite{Shankar2022}. AI safety and red-teaming characterize how systems can fail, informing the root-cause taxonomy forensics must apply but stop short of the reconstructive and evidentiary task. Forensics is the discipline that takes the artifacts and failure modes these fields describe and asks, after the fact and under constrained access, what happened and whether the answer will hold up.
    
\section{A Forensic Taxonomy of AI Systems}
\label{sec:taxonomy}
    The organizing claim of this paper is that the forensic character of an AI investigation is determined first by \textit{what the investigator can reach}. We therefore take investigator access as the primary axis of the taxonomy and treat system type as a secondary, annotative one. This section defines the access regimes, argues for their primacy, and presents the process-model matrix that the remainder of the paper elaborates.
    
    \subsection{The Primary Axis: Investigator Access}
    We distinguish three access regimes, defined by the artifacts an investigator can obtain rather than by the investigator's institutional role, since the same party may hold different access in different incidents.
    
    \textbf{White-box} access denotes possession of the model's internals and its provenance: trained weights and checkpoints, hyperparameters and training configuration, the training and fine-tuning data, and the pipeline that produced them. The canonical white-box investigator is the model owner examining its own system internally. At this access level, the full set of internal techniques is in principle available. These include attribution of training data, weight differencing, activation analysis, and the back door and watermark inspection, subject to scale and tooling limits.
    
    \textbf{Grey-box} access denotes possession of the deployment surface but not the model's internals: serving and inference logs, the prompts and system prompts in force, retrieval-augmented-generation (RAG) stores, agent action and tool-call traces, and orchestration configuration without the base model's weights or training data. The canonical grey-box investigator is an enterprise that deploys a third-party model. This is the most common posture in practice and the one in which contemporary incidents are most often first detected.
    
    \textbf{Black-box} access denotes possession only of inputs, outputs, and externally observable behavior; any deeper reach requires the cooperation of the party that holds the model. The canonical black-box investigator is an external regulator or incident-response team brought in after the fact. Here, the investigator must often reason about the system through surrogates and behavioral probing, and conclusions carry correspondingly lower evidentiary confidence.
    
    These regimes are best read as idealized access postures rather than rigid categories. In practice, access is often fragmented: an investigator may have weights but not training data, logs but not system prompts, or API transcripts but not the retrieval store. We therefore use white-, grey-, and black-box access as a coarse procedural shorthand. The value of the taxonomy is not the taxonomic purity but its link to what is forensically accessible in a given case.

    \subsection{Why Access Should Organize the Process Model}
    It is tempting to organize AI forensics by system type, since a Large Language Model (LLM), a tabular classifier, and a multi-agent system differ markedly in their artifacts. We instead use access to organize the process model because access determines what an investigator can collect, preserve, analyze, and report in a given case. System type remains important, but it does not by itself tell us what evidence is reachable.

    Our motivating incident makes this concrete. The financial advisory agent that begins to guide clients toward a fraudulent product is a system and an incident, but the three investigators face different forensic problems. The deploying firm, in a grey-box posture, can correlate logs, prompts, RAG state, and agent traces, and can replay against the components it controls, but cannot inspect the model that produced the behavior. The model provider, in a white box posture, can test whether the behaviour is linked to a poisoned fine tuning set using internal evidence that is unavailable to the others. The external regulator, in a black-box posture, observes only the harmful outputs and must reconstruct the cause through surrogate modeling, reaching a conclusion of materially lower confidence. The system type is fixed across all three; the access is not, and it is access that dictates what each can collect, preserve, conclude, and defend. This is the pattern made explicit in the process model matrix introduced in Section \ref{sec:process_model}.

    \subsection{The Secondary Axis: System Type}
    We do not argue that system type is unimportant. An LLM, a tabular classifier, and an agentic RAG system clearly produce different artifacts and failure modes. Our point is that access determines which of those artifacts an investigator can actually reach and use as evidence. System type shapes the evidence surface; access shapes the feasible investigation. We therefore use access as the main organizing axis, while treating system type as a secondary qualifier.

    System type remains forensically relevant and we retain it, but as an annotation rather than as structure. We order types by increasing forensic complexity: classical machine learning, deep neural networks, foundation models and LLMs, generative systems, and agentic or multi-agent systems. Moving along this axis enlarges the evidence surface. For example, agentic systems add tool call traces, inter agent messages, and autonomous scheduling state that simpler models lack. It also sharpens specific challenges, including nondeterminism and the volatility of intermediate state. We therefore use system type to qualify cells of the matrix where it materially changes what is achievable, without letting it fragment the process model.

\section{The AI Forensic Process Model}
\label{sec:process_model}

    We now develop the process model that follows from the access taxonomy. Table~\ref{tab:matrix} presents the central artifact of the paper. Its rows are the four forensic phases: collection, preservation, analysis, and reporting. Its columns are the three access regimes. Each cell states what is forensically achievable at that intersection, what is volatile or unreachable, and which existing work occupies it where such work exists. Reading down a column, the matrix gives a process for an investigator at a given access level. Reading across a row, it shows how the same forensic task changes as access narrows.

    The table also exposes weakly developed areas. Black box preservation, surrogate based black box analysis with quantified uncertainty, and chain of custody across organizational boundaries are still sparsely addressed. We mark these as gaps in Table~\ref{tab:matrix} and return to them in Section~\ref{sec:concluding_remarks}, where they form the basis of the research agenda.
    
    The following subsections explain the four phases in turn. Each phase is examined across white box, grey box, and black box access, and the motivating incident is used to show how the same event leads to different evidence and different levels of confidence.

    \begin{table}[!t]
        \centering
        \caption{The AI forensic process model as a function of investigator access. Each cell states what is forensically achievable, the binding challenge, and the prior work occupying it; italicised \emph{Gap} entries mark cells the         literature leaves thin and feed the research agenda (Sect.~\ref{sec:concluding_remarks}).
        }
        \label{tab:matrix}
        \footnotesize
        \setlength{\tabcolsep}{4pt}
        \renewcommand{\arraystretch}{1.3}
        \begin{tabular}{@{}>{\raggedright\arraybackslash}p{1.9cm}
                          >{\raggedright\arraybackslash}p{3.25cm}
                          >{\raggedright\arraybackslash}p{3.25cm}
                          >{\raggedright\arraybackslash}p{3.25cm}@{}}
        \toprule
        \textbf{Phase} & \textbf{White-box} & \textbf{Grey-box} & \textbf{Black-box} \\
         & \scriptsize weights + training data + pipeline
         & \scriptsize logs/API + deployment artifacts
         & \scriptsize I/O only; provider-gated \\
        \midrule
        
        \textbf{Collection}
        & Full artifact capture: weights, checkpoints, hyperparameters, training and inference logs, training data~\cite{sremack2026,baggili2019}.
        & Serving/inference logs, prompts and system prompts, RAG store, agent and tool-call traces, orchestration config; no base weights or training data~\cite{openclaw2026}.
        & Inputs, outputs, timing, and observable behavior only; deeper reach requires provider cooperation. \emph{Gap: standardized provider evidence disclosure.} \\
        \addlinespace[2pt]
        
        \textbf{Preservation}
        & Hash and sign weights and datasets; snapshot checkpoints; record provenance and model lineage~\cite{modeldna2023}.
        & Snapshot mutable RAG/vector store and logs; freeze config versions. \emph{Challenge:} state mutates; non-determinism.
        & Preserve I/O transcripts, timestamps, API version. \emph{Challenge:} silent provider updates break reproducibility. \emph{Gap: model-version attestation.} \\
        \addlinespace[2pt]
        
        \textbf{Analysis}
        & Training-data attribution / influence functions; weight diffing; activation probing; backdoor and watermark checks; membership inference~\cite{koh2017,baggili2019}.
        & Log/trace correlation; prompt-injection and RAG-poisoning reconstruction; replay against reachable components; drift analysis~\cite{openclaw2026}.
        & Model extraction/inversion to a surrogate; behavioral probing; reanimation; output statistics~\cite{aip2024}. \emph{Gap: uncertainty quantification for surrogate findings.} \\
        \addlinespace[2pt]
        
        \textbf{Reporting}
        & High-fidelity causal report; reproducible findings. \emph{Challenge:} explainability to court.
        & Probabilistic attribution; partial reconstruction; regulatory reporting (EU AI Act Art.~73).
        & Low-confidence behavioral findings; admissibility hardest. \emph{Gap: evidentiary standards and uncertainty communication.} \\
        
        \bottomrule
        \end{tabular}
    \end{table}

    \subsection{Collection}
    Collection is the identification and acquisition of evidence related to an incident. For AI systems, the candidate evidence is unusually heterogeneous, spanning the model's construction, its deployment, and its runtime. A reasonably complete inventory includes: training and fine-tuning data together with their provenance; weights and checkpoints; hyperparameters and training configuration; training logs; serving and inference logs; the prompts and system prompts in force at the time; retrieval-augmented-generation (RAG) and other vector-store state; agent action and tool-call traces; orchestration and routing configuration; accelerator memory holding activations and the key–value (KV) cache; the assembled context window presented to the model at inference; and in-flight network input/output. No single investigator typically commands all of these, which is precisely why the access stratifies the phase.
    
    A white-box investigator can capture all artifacts, acquiring the internals of the model and their provenance alongside the deployment and runtime evidence (Table~\ref{tab:matrix}, collection/white)~\cite{sremack2026}. A grey-box investigator collects the deployment surface — serving and inference logs, prompts and system prompts, RAG state, agent and tool-call traces, and orchestration configuration — but cannot reach the base model's weights or training data; this is the cell into which the agentic artifact planes of previous work fall most naturally~\cite{openclaw2026}. A black-box investigator is restricted to inputs, outputs, timing, and externally observable behavior, and must obtain anything further through provider cooperation, for which no standardized disclosure mechanism exists yet (Table~\ref{tab:matrix}, collection/black).
    
    Collection is further complicated by volatility: much of the most proximate evidence — activations, the KV cache, the assembled context window, in-flight I/O — exists only fleetingly and is rarely persisted, while the more durable artifacts may be the least reachable. The sequencing of acquisition is therefore not arbitrary but is governed by an order of volatility specific to AI systems, which we set out in Section~\ref{sec:volatility}. The practical rule is to collect the most volatile reachable evidence first, before it decays, regardless of how probative the durable artifacts would be if they could be obtained.
    
    \paragraph{Case beat.}
    In our incident, the deploying firm can collect serving logs, prompts, customer interaction history, the retained RAG store, tool call traces, and policy check logs, but not the model itself. The provider can additionally capture the deployed weights, fine tuning data, and training configuration. The regulator initially has only reported outputs and whatever the firm provides, and must request further evidence. The same incident therefore yields three different collectible sets.
    
    \subsection{Preservation and Chain of Custody}
    Preservation is the maintenance of collected evidence in a state whose integrity can be demonstrated, together with a documented chain of custody establishing who handled it and how. For conventional evidence, this is well understood: hash on acquisition, store immutably, log every transfer. AI evidence admits the same techniques for its static artifacts — weights and datasets can be cryptographically hashed and signed, checkpoints can be captured and provenance recorded through data sheets, model cards, and an AI bill of materials (AIBOM); fingerprinting methods such as model-DNA support claims about whether one model derives from another (Table~\ref{tab:matrix}, preservation/white)~\cite{modeldna2023}. The difficulty lies with everything that is not static.
    
    Two problems are distinctive. First, much forensic state is mutable and must be snapshotted at a known instant or it is lost: a RAG store is updated as documents change, an agent's memory accrues, and configuration is rolled forward. A grey-box investigator must freeze these — versioned snapshots of the vector store, the logs, and the configuration in force — accepting that what is preserved is a point-in-time view of a moving target (Table~\ref{tab:matrix}, preservation/grey). Second, and more fundamentally, AI behavior is non-deterministic, so even faithfully preserved evidence may not reproduce on replay; and in the black-box case the provider may silently update or retire the model between the incident and the investigation, so that the very system under examination no longer exists in the form that produced the harm.
    
    This last point exposes a missing forensic primitive. Conventional forensics can often reexamine a preserved disk because the investigator holds a stable copy. AI forensics is different when the investigator does not hold the model and the provider may have changed it after the incident. We therefore identify \emph{model version attestation} as a needed preservation capability. By this we mean a verifiable and timestamped binding between an observed behavior and the specific model version, configuration, retrieval state, and serving context that produced it (Table~\ref{tab:matrix}, preservation/black). Without it, black-box findings rest on the assumption that the model examined is the model that produced the harmful behaviour, an assumption the investigator usually cannot verify.
    
    \paragraph{Case beat.}
    The provider snapshots and signs the deployed weights and fine tuning data. The firm freezes the RAG store and exports traces with hashes and timestamps, preserving a grey box view of the deployment at the time of preservation. The regulator can preserve only transcripts and reported version strings, and later learns that the model has already been updated. The absence of model version attestation therefore becomes concrete.

    \subsection{Analysis and Attribution}
    Analysis interprets preserved evidence to determine what happened and why. Attribution then links the behavior to a likely cause. We organize the possible causes into a root cause taxonomy that includes data poisoning, backdoor or trojan insertion, prompt injection, RAG poisoning, jailbreak, model drift, and supply chain or model substitution compromise. This taxonomy should also include the null hypothesis of benign but harmful behavior, where a faithfully functioning model produces a harmful output without an attack. Framing the causes in terms familiar to the security community connects forensic findings to known attack techniques without forcing the investigator to assume that every incident is adversarial.

    The techniques available to establish a cause depend strongly on access (Table~\ref{tab:matrix}). A white box investigator can use internal evidence such as training data attribution, influence based analysis, weight comparison against a known good baseline, activation probing, backdoor checks, watermark or canary inspection, and membership inference~\cite{koh2017,baggili2019}. These techniques should be treated as forensic indicators rather than automatic proof of causality. In large models, they are often approximate, sensitive to assumptions, and difficult to reproduce independently. A grey box investigator works mainly through correlation and reconstruction, by aligning logs and traces, reconstructing a prompt injection or RAG poisoning pathway from the recorded context, replaying inputs against reachable components, and testing for drift~\cite{openclaw2026}. A black box investigator, lacking the model, must rely on behavioral investigation, output statistics, surrogate modeling, model extraction or inversion, and reanimation style reconstruction~\cite{tramer2016,aip2024}. The central difficulty is to separate causes that may look similar from the outside, including model resident failures, input driven failures, retrieval poisoning, and supply chain compromise.

    \paragraph{Case beat.}
    The provider uses training data attribution to link the steering behaviour to poisoned fine tuning examples, supporting a high confidence white box attribution. The firm correlates prompts, RAG entries, tool traces, and policy logs, showing that the RAG entry alone does not reproduce the behaviour and that the recorded policy checks executed as configured. This supports a medium confidence grey box finding. The regulator can show systematic harmful behaviour through a surrogate, but cannot independently separate model resident from input resident causes. The result is a low confidence black box conclusion.

    \subsection{Reporting}
    Reporting communicates findings to the parties entitled to act on them, including internal decision makers, regulators, or a court. This is where forensic work meets evidentiary and legal standards. Four demands dominate: admissibility, requiring that the method and custody withstand challenge; explainability, requiring that a non-specialist audience can follow how a conclusion was reached; faithful communication of uncertainty, so that confidence is neither overstated nor lost; and compliance with regulatory reporting obligations and their deadlines.
    
    Reporting inherits the access stratification of the phases before it: confidence degrades from white through grey to black, and admissibility is hardest precisely where access is lowest (Table~\ref{tab:matrix}, reporting row). A white-box report can offer a high-fidelity causal account with reproducible findings, its remaining challenge being to render internal techniques such as influence-function attribution intelligible and credible to a court. A grey-box report offers probabilistic attribution and partial reconstruction, and is typically the form in which a regulatory filing is made. A black box report must present behavioral findings with low confidence, and its admissibility is often the most fragile. The field also lacks agreed evidentiary standards and a clear vocabulary to communicate uncertainty in such cases. We return to this gap in Section~\ref{sec:concluding_remarks}.

    \paragraph{Case beat.}
    The firm files a serious incident report using its grey box reconstruction and the provider's white box attribution where cooperation is available. The report grades each claim by access: the poisoned fine tune as high confidence, the deployment pathway as medium confidence, and the externally observed behaviour as low confidence because the model has since changed.

\section{An Order of Volatility for AI Systems}
\label{sec:volatility}
    Traditional digital forensics often follows an order of volatility, as described in RFC 3227 \cite{rfc3227}, which advises investigators to collect the most ephemeral evidence first. The usual sequence starts with processor state and memory, then moves toward disk, removable media, and archival storage. AI systems require their own ordering because their evidence hierarchy does not map onto storage media but onto the stages of an inference and training pipeline, with very different decay rates. We propose the following order, from most to least volatile, as an AI specific order inspired by RFC 3227, and use it to dictate collection sequencing (Section~\ref{sec:process_model}).

    \begin{enumerate}
        \item Accelerator memory. This includes activations, the key value cache, and in flight inference state resident on the GPU or TPU. It is the most volatile layer because it may be overwritten within or between requests. With current tooling, recovery is largely aspirational and usually requires instrumentation prior to the incident rather than acquisition after it.
        \item The assembled context window. This includes the full prompt, system prompt, and retrieved content actually presented to the model at inference. It can be recovered if it is logged or intercepted at the serving layer. If it is not persisted, it is usually lost with the request.

        \item In flight network and tool call I/O. This includes requests, responses, and the agent's outbound tool invocations while they are in transit. It can be recovered through network capture and proxy techniques, provided that capture is active at the time.
        
        \item Host memory and orchestration runtime state. This includes the serving process and agent framework state in system RAM. It is recoverable in principle through conventional memory forensics, although parsing AI framework structures from a memory image is not yet well supported.
        
        \item Serving and inference logs. This includes request and response logs, routing decisions, and framework traces, if logging is enabled. When present, these logs are usually recoverable. The main constraint is often whether they exist at all, rather than whether they can be acquired. Regulation that mandates logging, such as the EU AI Act provisions discussed in Section~\ref{sec:regulatory}, acts directly on this layer.

        \item RAG and vector store state. This includes the retrieval corpus and its embeddings. It is recoverable but mutable. The store changes as documents are added, removed, or re indexed, so a forensic copy is a point in time snapshot whose value may decay as the live store moves on.

        \item Agent and tool call traces and agent memory. This includes persisted records of an agent's actions and accumulated memory. It is recoverable where the orchestration layer records it, which is increasingly common but not guaranteed.
        
        \item Model checkpoints and weights. This includes the trained parameters. These artifacts are relatively stable and straightforward to recover for those who hold them, but inaccessible to investigators who do not.
        
        \item Training data, pipeline, and lineage. This includes the data and process that produced the model, together with its provenance. It is the most stable layer and often the most probative for questions of poisoning or origin, but it is also often the least reachable because it usually remains with the original developer.

    \end{enumerate}

    \paragraph{The volatility reachability inversion.}
    The payoff of this ordering is the structural tension between how quickly evidence decays and who can realistically reach it. The most volatile evidence, such as accelerator state, the assembled context window, and in flight tool or network activity, is often available only if logging or instrumentation were already in place. A black box investigator may see only the external I/O and timing, while a grey box investigator may reach logs, traces, RAG state, and orchestration state if these were retained. The most stable and often most probative evidence, including weights, checkpoints, training data, and lineage, usually remains with the provider. The evidence that can best explain the cause of an incident is therefore often the evidence least reachable to the party first affected by it. This inversion motivates the access conditioned examination framework of Section~\ref{sec:framework} and the research agenda of Section~\ref{sec:concluding_remarks}.

\section{Challenges}
\label{sec:challenges}
    The preceding sections surface a set of obstacles that recur across phases and access levels. We collect them here as standing challenges of AI forensics; several are discussed as research directions in Section~\ref{sec:concluding_remarks}.

    \paragraph{Non determinism.}
    AI behaviour is not always reproducible from the input alone. Replay may depend on model version, preprocessing, inference settings, random seed, retrieval state, tool state, model routing, and provider updates. Investigators therefore need to preserve execution context and state conclusions with appropriate uncertainty.
    
    \paragraph{Volatility and scale.}
    The most proximate evidence may decay within or between requests, while durable evidence may consist of billions of parameters and large training corpora. AI forensics must therefore combine rapid capture of ephemeral state with scalable methods for preserving and analysing large artifacts.
    
    \paragraph{Black box opacity and provider non cooperation.}
    The party that detects harm often cannot reach the model that produced it. In the absence of standard disclosure mechanisms, black box investigations depend on provider cooperation, model version attestation, and careful reporting of what could not be verified.
    
    \paragraph{AI specific anti forensics.}
    Adversaries may design backdoors that evade inspection, poisoning that resists attribution, or logging practices that suppress useful traces. Non determinism can also provide plausible deniability, making it harder to distinguish attack, drift, and benign but harmful behaviour.
    
    \paragraph{Standards, legal process, and confidentiality.}
    The field lacks mature standards, tools, and trained investigators comparable to those in disk or memory forensics. It also faces cross border legal questions because training, hosting, deployment, evidence, and parties may sit in different jurisdictions. Finally, the most probative evidence may contain personal data or proprietary assets, so forensic access must be balanced against privacy and confidentiality constraints.

\section{The Regulatory Driver}
\label{sec:regulatory}
    The discipline we have described is not just intellectually overdue; it is becoming a legal necessity. The EU AI Act applies progressively, and its obligations do not enter into force for all systems at the same time. Still, several of its requirements already point toward the same basic need: when a serious AI incident occurs, the relevant parties must be able to reconstruct what happened, identify the state of the system, and support their findings with records that can be examined.
    
    Most directly, the Act creates several preconditions for AI forensics. Article~12 requires high risk AI systems to support automatic event recording over their lifetime, with traceability sufficient to identify risk situations and reconstruct the system's functioning. Article~18 requires technical documentation, and Article~19 concerns the retention of automatically generated logs. Article~73 places serious incident reporting duties on providers of high risk AI systems, while Article~26 requires deployers that identify a serious incident to inform the provider and the relevant authorities; where the provider cannot be reached, Article~73 applies to the deployer mutatis mutandis~\cite{euaiact2024}. In the terms of Section~\ref{sec:volatility}, regulation acts directly on the log layer of the volatility order and raises the floor on what a grey box investigator can expect to find. It also points toward the need for model version attestation, although the Act does not yet define such attestation as a concrete forensic mechanism.
    
    Regulation is, however, a driver rather than a ceiling, and it would be a mistake to scope AI forensics to compliance. The demand for forensic analysis of models predates the Act and extends well beyond it: into mergers and acquisitions, restructuring, private-equity due diligence, vendor assurance, and litigation, where a model's behavior or provenance must be established to an evidentiary standard~\cite{am2025}. The EU AI Act is best understood as a forcing function that makes the discipline urgent and gives it a first well defined field of application. It should not be treated as the boundary of AI forensics. The methods, process model, and order of volatility we propose are intended to serve a wider need, with regulatory compliance as one important instance.

\section{A Practical Examination Framework}
\label{sec:framework}

\begin{figure}[h]
        \centering
        \includegraphics[width=\textwidth]{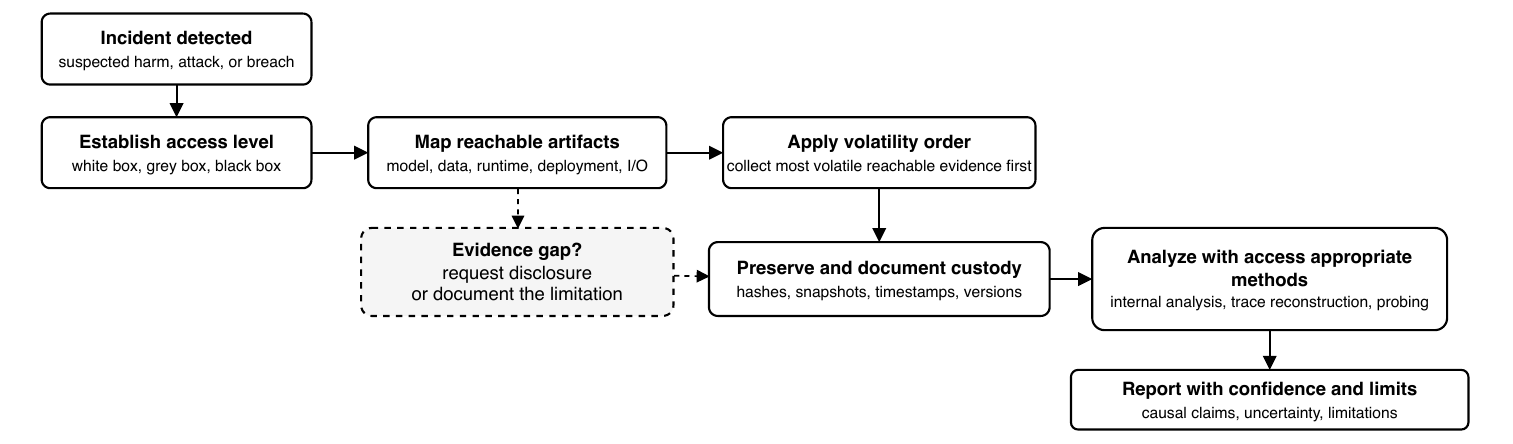}
        \caption{Access conditioned workflow for AI forensic investigation.}
        \label{fig:workflow}
    \end{figure}

    The contributions so far describe what is achievable at each intersection of phase and access. We now turn the matrix into a practical workflow. As shown in Figure~\ref{fig:workflow}, an AI forensic investigation should first establish the investigator's access level. This determines which artifacts are reachable and which claims can later be supported. The investigator then maps the reachable artifacts, applies the order of volatility to decide what must be collected first, preserves the evidence with custody records, analyzes the evidence using methods supported by the access level, and reports the findings with explicit confidence and limitations. If the available evidence is insufficient, the investigator should request provider disclosure where possible or document the access gap as part of the report. 

    The three playbooks below do not redefine the access regimes. Instead, they translate the columns of Table~\ref{tab:matrix} into practical steps, using the order of volatility in Section~\ref{sec:volatility} to decide what should be collected first.

    \paragraph{White box.}
    In a white box investigation, the investigator should begin with volatile runtime evidence before moving to durable artifacts such as weights, checkpoints, training data, fine tuning data, and configuration. Preservation should hash and sign weights and datasets, snapshot checkpoints, and record provenance through datasheets, model cards, and an AIBOM. Analysis can use internal evidence, including training data attribution, weight comparison, activation inspection, and backdoor checks. The report can support stronger causal claims, but should still explain the methods and their limits in terms that a non specialist audience can assess.

    \paragraph{Grey box.}
    In a grey box investigation, the priority is to preserve the deployment surface before it changes. This includes the assembled context, serving logs, prompts, RAG state, agent traces, tool call records, and configuration. Preservation should freeze versioned snapshots of the store, logs, and configuration, with hashes and timestamps where possible. Analysis is mainly reconstructive: the investigator correlates logs and traces, tests possible prompt injection or retrieval poisoning paths, and replays inputs against reachable components. The report should present partial attribution, explain which model internals were unavailable, and avoid implying a stronger causal claim than the evidence supports.

    \paragraph{Black box.}
    In a black box investigation, the investigator can usually preserve only external inputs, outputs, timing, and reported model or API version information. Provider disclosure should be requested early, since later model updates may weaken reproducibility. Preservation should keep I/O transcripts with timestamps and any available attestation records. Analysis relies on behavioral probing, output statistics, surrogate modelling, and reconstructed evidence where available. The report should treat these findings as behavioral rather than fully causal, with explicit uncertainty and clear limits.

   We offer this workflow as a first scaffold, not a validated standard. It has not yet been evaluated across a corpus of real investigations, and some steps, especially black box preservation, depend on capabilities that the field has not yet matured. Its value is to make the proposed structure actionable and to give the community a concrete object to test, challenge, and refine.

\section{Concluding Remarks and Future Work}
\label{sec:concluding_remarks}

    This paper started from a simple observation. AI forensics, understood as the forensic investigation of AI systems, already has a name, but the literature around it remains fragmented. Existing work has examined model forensics, agentic system forensics, reconstruction tooling, provenance methods, and practitioner needs, often with different assumptions about system type and investigator access. We have argued that these efforts can be read as parts of one discipline if they are organized around what the investigator can actually reach. The process model matrix in Table~\ref{tab:matrix} makes this argument concrete. It places existing work within a common structure and also shows where the structure is still weak. From these gaps we derived an AI specific order of volatility and an access conditioned workflow that turns the matrix into a practical starting point for investigation.
    
   Black box preservation remains especially underdeveloped. It depends on model version attestation, meaning a verifiable and timestamped link between an observed behavior and the model version, configuration, retrieval state, and serving context that produced it. Without such a capability, black box findings may rest on assumptions that the investigator cannot verify. Surrogate based analysis raises a related problem. When the system being examined is only an approximation of the system that produced the incident, investigators need principled ways to express confidence and uncertainty. Provider evidence disclosure is another open problem, since grey box and black box investigators often depend on voluntary cooperation to reach durable evidence held by the provider. Finally, chain of custody for mutable and nondeterministic artifacts remains unresolved. RAG stores may drift, agent memory may change, and behavior may not reproduce even when the available evidence has been preserved.
    
    Forensic readiness by design is important, but it should be treated as one part of the agenda rather than as a complete solution. Durable logging, provenance capture, attestable versioning, and preserved execution context would improve nearly every cell of the matrix. Regulation, including the EU AI Act's record keeping provisions, already pushes in this direction (Section~\ref{sec:regulatory}). The harder problem is adoption. The challenge is to make forensic readiness cheap, default, and practical enough to be used, rather than merely recommended.

    We do not claim to settle AI forensics. The field is still young, and a position paper can only provide a structure for further work. Our aim has been to offer an organizing axis, a process model matrix, an order of volatility, and a practical workflow. These are meant to help the community locate existing work, identify missing capabilities, and direct effort toward underdeveloped parts of the field.

\begin{credits}
\subsubsection{\ackname} This research was supported by the Natural Sciences and Engineering Research Council of Canada (NSERC) through the NSERC-CSE Research Community Grants (ALLRP 598786-24), the Discovery Grants Program (RGPIN-2026-04369), the Canada Research Chairs Program (CRC-2024-00017), and the CREATE Program (CREATE 596346-2025). The views, findings, and conclusions expressed here are solely those of the authors and do not necessarily reflect the official policies or positions of the Communications Security Establishment (CSE) or the Government of Canada.  

\subsubsection{\discintname}
The authors declare no competing interests.
\end{credits}

\bibliographystyle{splncs04}
\bibliography{bib}

\end{document}